\documentclass[12pt, a4paper]{article}
\usepackage{cite}
\usepackage{amsmath,amssymb}
\input{colordvi.tex}
\usepackage{comment}
\usepackage{bm}
\usepackage{url}
\usepackage{ifpdf}
\ifpdf
  \usepackage{graphicx, hyperref, xcolor}     
\else     
  \usepackage[dvipdfmx]{graphicx, hyperref, xcolor}     
 \fi

\definecolor{rossoferrari}{HTML}{D9073D}
\definecolor{mediumblue}{HTML}{0000CD}
\hypersetup{
setpagesize=false,
bookmarksnumbered=true,%
bookmarksopen=true,%
colorlinks=true,%
linkcolor=rossoferrari,
urlcolor=mediumblue,
citecolor=mediumblue,
}

\begin{document}


\begin{center}

\begin{flushright}
TU-1136
\end{flushright}

\vskip .75in

{\Large \bf
Electric current on surface of a metal/superconductor \\[.3em] in axion/hidden-photon background
}

\vskip .75in

{\large Yasuhiro Kishimoto$^{(a, b)}$ and Kazunori Nakayama$^{(c,d)}$}

\vskip 0.25in
$^{(a)}${\em Research Center for Neutrino Science, Tohoku University,  Sendai, Miyagi 980-0845, Japan}\\
$^{(b)}${\em Kavli IPMU (WPI), The University of Tokyo,  Kashiwa, Chiba 277-8583, Japan}
$^{(c)}${\em Department of Physics, Tohoku University, Sendai, Miyagi 980-8578, Japan}
$^{(d)}${\em Department of Physics, Faculty of Science,\\ The University of Tokyo,  Bunkyo-ku, Tokyo 113-0033, Japan}\\[.3em]

\end{center}
\vskip .5in

\begin{abstract}
We derive the electric current on surface of a normal metal or superconductor induced by the light bosonic dark matter, such as axion or hidden photon, by appropriately taking account of the boundary condition of the electric and magnetic field at the surface. We discuss detection possibility of such an electric current.
\end{abstract}



\renewcommand{\thepage}{\arabic{page}}
\setcounter{page}{1}
\renewcommand{\thefootnote}{\#\arabic{footnote}}
\setcounter{footnote}{0}

\newpage

\tableofcontents

\section{Introduction} \label{sec:Intro}

The QCD axion or the axion-like particle is one of the best-motivated candidates of light dark matter (DM) in the universe~\cite{Preskill:1982cy,Abbott:1982af,Dine:1982ah,Kim:1986ax,Kawasaki:2013ae}.
Another well-motivated light DM candidate is the hidden photon. Several production mechanisms of the hidden photon to explain the observed DM abundance have been proposed~\cite{Agrawal:2018vin,Co:2018lka,Bastero-Gil:2018uel,Dror:2018pdh,Long:2019lwl,Graham:2015rva,Ema:2019yrd,Ahmed:2020fhc,Kolb:2020fwh,Nakayama:2021avl}.
There are many on-going and proposed experiments to probe the axion-like particle or hidden photon DM, although it is not discovered yet~\cite{Sikivie:1983ip,Bradley:2003kg,Asztalos:2009yp,Zhong:2018rsr,McAllister:2017lkb,Alesini:2017ifp,Semertzidis:2019gkj,Horns:2012jf,Jaeckel:2013eha,TheMADMAXWorkingGroup:2016hpc,Sikivie:2013laa,Chaudhuri:2014dla,Kahn:2016aff,Obata:2018vvr,Nagano:2019rbw,Lawson:2019brd,Zarei:2019sva,Budker:2013hfa,Barbieri:1985cp,Barbieri:2016vwg,Chigusa:2020gfs,Ikeda:2021mlv,Marsh:2018dlj,Schutte-Engel:2021bqm,Chigusa:2021mci,Crescini:2020cvl,Hochberg:2015pha,Hochberg:2016ajh,Hochberg:2016sqx,Hochberg:2017wce,Knapen:2017ekk,Griffin:2018bjn,Hochberg:2019cyy,Campbell-Deem:2019hdx,Mitridate:2020kly,Hochberg:2021yud,Iwazaki:2020agl,Iwazaki:2020zer}.

Recently, Refs.~\cite{Iwazaki:2020agl,Iwazaki:2020zer} proposed a novel detection method of axion DM. Under the axion DM background and the applied external magnetic field, the oscillating electric field is induced, which leads to oscillating electric current on a superconductor or normal metal. Although the magnetic field is screened inside the superconductor, there is a thin layer region around its surface in which the external magnetic field exists. Then the oscillating current emits the dipole radiation and it can be detected by radio antenna. 
The same situation in the case of perfect metal was analyzed in Ref.~\cite{Horns:2012jf} and there it was found that the electric field vanishes at the surface. Thus there should be no surface electric current in such a case. On the other hand, Ref.~\cite{Iwazaki:2020zer} analyzed the case of a metal with finite conductivity and claimed that the electric current is induced in a thin layer around a metal surface.

In this paper we revisit this idea by appropriately taking account of the boundary condition at the surface of a metal/superconductor. We also consider not only the case of axion DM but also the hidden photon DM as an extension of the previous analysis.
We find that the current on a surface of metal or superconductor is orders-of-magnitude smaller than the estimate of Refs.~\cite{Iwazaki:2020agl,Iwazaki:2020zer}. 
Still we find a substantial electric current on the surface and derive a configuration of oscillating electric field outside the metal/superconductor. This may be partly relevant to the DM detection ideas using mirrors such as MADMAX or BRASS~\cite{Horns:2012jf,TheMADMAXWorkingGroup:2016hpc} and also for cavity experiments such as ADMX~\cite{Asztalos:2009yp}, HAYSTAC~\cite{Zhong:2018rsr}, ORGAN~\cite{McAllister:2017lkb}, KLASH~\cite{Alesini:2017ifp} and CULTASK~\cite{Semertzidis:2019gkj}.

In Sec.~\ref{sec:ed} we briefly review the electrodynamics in the metal and superconductor. In Sec.~\ref{sec:metaldm} we study the effect of axion and hidden photon DM on the electric field configuration around the metal surface and derive the electric current on the surface of the metal. In Sec.~\ref{sec:scdm} we extend the analysis to the case of superconductor. We discuss possible detection strategy of such DM-induced currents in Sec.~\ref{sec:det}. Sec.~\ref{sec:dis} is devoted to summary.

\section{Electrodynamics of metal/superconductor}  \label{sec:ed}

In this section we briefly summarize the basic equations of electric/magnetic field in the metal and superconductor for later use.
In this paper we take the dielectric constant $\epsilon$ and permeability $\mu$ of the material to be equal to unity for simplicity. This assumption does not much affect the result as far as $|\epsilon-1|, |\mu-1| \lesssim \sqrt{m/\sigma}$ or $m/m_A$ for a normal metal or superconductor respectively, where $m$ is the DM mass, $\sigma$ is the conductivity and $m_A$ is the photon mass in the superconductor.

\subsection{Metal}

\subsubsection{Maxwell equations}

Let us consider the Lagrangian:
\begin{align}
	\mathcal L = -\frac{1}{4}F_{\mu\nu}F^{\mu\nu} - e A_\mu j^\mu.
\end{align}
where $F_{\mu\nu} = \partial_\mu A_\nu-\partial_\nu A_\mu$ is the electromagnetic field strength tensor, $e$ is the gauge coupling constant (or $e=\sqrt{4\pi\alpha_e}$ with $\alpha_e$ being the fine structure constant) and $j_\mu$ is the electric current.
The equation of motion (Maxwell equation) is given by
\begin{align}
	\partial_\mu F^{\mu\nu} = \Box A^\nu - \partial^\nu (\partial_\mu A^\mu) = ej^\nu.
\end{align}
Applying $\partial_\nu$ on both sides leads to the current conservation: $\partial_\nu j^\nu=0$.
In components, the equation of motion leads to
\begin{align}
\begin{cases}
	&\nu=0~~:~~\Delta A^0 + \vec\nabla\cdot\dot{\vec A}= ej^0\\
	&\nu=i~~:~~\Box \vec A -\vec \nabla(\dot A^0 + \vec\nabla\cdot\vec A) = e\vec j
\end{cases}
\end{align}
These equations are rewritten in terms of $\vec E=-\vec\nabla A^0 - \dot{\vec A}$ and $\vec B=\vec\nabla\times \vec A$:
\begin{align}
\begin{cases}
	&\nu=0~~:~~-\vec\nabla\cdot\vec E= ej^0\\
	&\nu=i~~:~~\dot{\vec E}-\vec\nabla\times\vec B = e\vec j
\end{cases}
\end{align}
From this we obtain
\begin{align}
	&\Box \vec E = -e(\dot{\vec j} + \vec\nabla j^0), \label{E_eq}  \\
	&\Box \vec B = e \vec\nabla\times\vec j \label{B_eq}.
\end{align}
On the other hand, there are identities:
\begin{align}
	\vec\nabla\cdot\vec B=0,~~~~~~~\vec\nabla\times\vec E= -\dot{\vec B}.
\end{align}

\subsubsection{Plasma frequency}   \label{sec:plasma}
 
 Let us express the current $\vec j$ in terms of the velocity of free electrons in the metal.\footnote{
 	In our convention, $\vec j$ represents the electron number current and the electric current is given by $-e\vec j$.
 }
 The equation of motion of free electron under the electric field is given by
\begin{align}
	m_e \dot{\vec v}_e = -e\vec E,
\end{align}
where $m_e$ is the electron mass. Thus the current is given by
\begin{align}
	\dot{\vec j} = n_e \dot{\vec v}_e = -\frac{en_e}{m_e} \vec E,  \label{dotj}
\end{align}
where $n_e$ is the electron number density.
Substituting this into (\ref{E_eq}) and assuming the spatial gradient $\vec\nabla j^0$ is small enough, we obtain
\begin{align}
	\Box \vec E = \frac{e^2n_e}{m_e} \vec E \equiv \omega_p^2 \vec E.
\end{align}
Here we have defined the plasma frequency $\omega_p$. It shows that the electric field obtains an effective plasma mass $\omega_p^2$.
From (\ref{dotj}) we also obtain
\begin{align}
	\frac{d}{dt}\left(  \vec\nabla\times\vec j - \frac{en_e}{m_e}\vec B \right)=0.
\end{align}
It is solved as
\begin{align}
	 \vec\nabla\times\vec j = \frac{en_e}{m_e}(\vec B - \vec B_0),  \label{London_metal}
\end{align}
where $\vec B_0$ represents arbitrary constant. When $\vec B_0=0$, this is called the London equation as shown in the next subsection.
Substituting this into (\ref{B_eq}), we obtain
\begin{align}
	\Box \vec B =\omega_p^2 (\vec B - \vec B_0).
\end{align}

To summarize, we obtained
\begin{align}
	&\Box \vec E = \omega_p^2 \vec E \\
	&\Box \vec B = \omega_p^2 (\vec B - \vec B_0).  
\end{align}
For the electric field, the static solution should be of the form
\begin{align}
	\vec E(x) = \vec E_0 e^{-\omega_p x}.
\end{align}
For the magnetic field, the constant solution is allowed:
\begin{align}
	\vec B(x) = \vec B_0.
\end{align}

\subsubsection{Finite conductivity} \label{sec:cond}

So far we have neglected the dissipation of the electron, i.e., we considered the limit of perfect metal. 
Several effects such as impurities in the material and scattering with phonons can act as an effective dissipation for the electron motion.
We can phenomenologically introduce the effect of dissipation as
\begin{align}
	\dot{\vec v}_e + \Gamma \vec v_e = -\frac{e \vec E}{m_e},
\end{align}
where $\Gamma$ denotes the dissipation coefficient.
By defining the Fourier transform as
\begin{align}
	\vec E(t) = \int \vec E_{\omega} e^{i\omega t} d\omega,~~~~~~
	\vec E_{\omega} = \frac{1}{2\pi}\int \vec E(t) e^{-i\omega t} dt,
\end{align}
and so on, we obtain
\begin{align}
	e\vec j_\omega= en_e \vec v_\omega = -\sigma \vec E_\omega,~~~~~~
	\sigma\equiv \frac{1}{i\omega + \Gamma}\frac{e^2 n_e}{m_e} = \frac{\omega_p^2}{i\omega + \Gamma}.
	\label{sigma}
\end{align}

For a typical metal, $\omega_p \sim 10\,{\rm eV}$ and $\Gamma \sim 0.1\,{\rm eV}$. 
Later we will consider $\omega$ equal to the DM mass, which is typically much smaller than $\Gamma$. For $\omega\ll \Gamma$, we have
\begin{align}
	e\vec j= -\sigma \vec E;~~~~~~ \sigma \simeq \frac{e^2n_e}{m_e \Gamma} = \frac{\omega_p^2}{\Gamma}.
\end{align}
Then Eq.~(\ref{E_eq}) and (\ref{B_eq}) becomes
\begin{align}
	&\Box \vec E -\sigma \dot{\vec E} = 0,\\
	&\Box \vec B -\sigma \dot{\vec B} = 0.
\end{align}
By assuming the solution of the form $e^{i(-\omega t+kx)}$, we obtain the dispersion relation as
\begin{align}
	k = (1+i)\delta(\omega)^{-1},~~~~~~\delta(\omega)\equiv \sqrt{\frac{2}{\sigma\omega}}.
	\label{skin}
\end{align}
Thus it describes the exponentially decaying solution $\propto e^{-x/\delta}$ with a skin depth $\delta$.

\subsection{Superconductor}  \label{sec:sc}

Let us consider the following Lagrangian:
\begin{align}
	\mathcal L &= -\frac{1}{4}F_{\mu\nu}F^{\mu\nu} - |(\partial_\mu- ieqA_\mu)\Phi|^2 - V(|\Phi|)\\
	&=-\frac{1}{4}F_{\mu\nu}F^{\mu\nu} -|\partial_\mu\Phi|^2-e^2q^2 |\Phi|^2A_\mu A^\mu
	-ieqA_\mu(\Phi^*\partial^\mu\Phi -\Phi\partial^\mu\Phi^*) - V(|\Phi|).
\end{align}
It is invariant under the gauge transformation $\Phi\to e^{iq\chi(x)}\Phi$ with $q$ being the charge of $\Phi$ and $A_\mu\to A_\mu+(\partial_\mu\chi)/e$.
Let us suppose that the Higgs field obtains a VEV $|\Phi| = v_\Phi$. Note that there is an ambiguity of the phase: $\Phi= v_\Phi e^{i\theta(x)}$.
Neglecting the dynamics of the radial mode of the Higgs, the Lagrangian is also expressed as
\begin{align}
	\mathcal L =-\frac{1}{4}F_{\mu\nu}F^{\mu\nu} -\frac{1}{2}m_A^2 A_\mu A^\mu
	-eA_\mu j_\theta^\mu,
\end{align}
where $m_A^2= 2e^2 q^2 v_\Phi^2$ and $j_\theta^\mu = iq(\Phi^*\partial^\mu\Phi -\Phi\partial^\mu\Phi^*)=-2qv_\Phi^2 \partial^\mu \theta$.
The equation of motion is given by
\begin{align}
	\partial_\mu F^{\mu\nu} = \Box A^\nu - \partial^\nu (\partial_\mu A^\mu) = m_A^2 A^\nu + ej_\theta^\nu.
\end{align}
One can also define the ``current'' $e j^\nu_A \equiv m_A^2 A^\nu$ so that the equation is rewritten as
\begin{align}
	\Box A^\nu - \partial^\nu (\partial_\mu A^\mu) = e(j_A^\nu +j_\theta^\nu).
\end{align}
Notice that the current $j_A^\nu$ and $j_\theta^\nu$ are not gauge invariant, but the sum $j_A^\nu +j_\theta^\nu$ is gauge invariant.
By using this gauge degree of freedom, one can take a unitary gauge $\theta(x)=0$ so that $j_\theta^\mu=0$.

As a result, we obtain the Lagrangian in the unitary gauge\footnote{
	The BCS theory implies $m_A^2 \simeq e^2 n_e/m_e$, which is the same expression as the plasma frequency of a metal.
}
\begin{align}
	\mathcal L = -\frac{1}{4}F_{\mu\nu}F^{\mu\nu} - \frac{1}{2} m_A^2 A_\mu A^\mu.
\end{align}
The equation of motion is given by
\begin{align}
	\partial_\mu F^{\mu\nu} = \Box A^\nu - \partial^\nu (\partial_\mu A^\mu) = m_A^2 A^\nu = ej_A^\nu.
\end{align}
These equations are rewritten in terms of $\vec E$ and $\vec B$:
\begin{align}
\begin{cases}
	&\nu=0~~:~~-\vec\nabla\cdot\vec E= ej_A^0\\
	&\nu=i~~:~~\dot{\vec E}-\vec\nabla\times\vec B = e\vec j_A
\end{cases}
\end{align}
From this we obtain
\begin{align}
	&(\Box -m_A^2) \vec E =0,  \\
	&(\Box-m_A^2) \vec B = 0.
\end{align}
The static solution looks like
\begin{align}
	\vec E(x) = \vec E_0 e^{-m_A x},~~~~~~\vec B(x) = \vec B_0 e^{-m_A x}.
\end{align}
Therefore, not only $\vec E$, but also $\vec B$ is screened in the superconductor. Note that we have
\begin{align}
	 e\vec\nabla\times\vec j_A = m_A^2 \vec B.
	 \label{London}
\end{align}
This is the London equation and it corresponds to the case of $\vec B_0=0$ in Eq.~(\ref{London_metal}).

\section{Metal with dark matter background}  \label{sec:metaldm}

Now let us consider the effects of DM on the electrodynamics of metal. First we assume axion DM and derive the modified Maxwell equations in the presence of axion. Next we will also consider the cases of hidden photon DM.

\subsection{Axion dark matter}  \label{sec:axionmetal}

\subsubsection{Basic equations}

We are interested in the axion DM with the axion mass range of $\mu{\rm eV} \lesssim m_a\lesssim {\rm meV}$. The number density of such a light bosonic DM around the Earth is huge and we can regard the axion DM as a coherent field within its de Broglie wavelength. We take the axion field as
\begin{align}
	a(\vec x,t) \simeq a_{0} \cos\left( m_at - m_a \vec v_a\cdot\vec x \right).
\end{align}
The axion velocity is about $|\vec v_a|\sim 10^{-3}$ around the Earth. In the most part of the following discussion, we neglect the spatial dependence of the axion field, which is justified as far as we consider the dynamics within one coherent time $\tau \sim 1/(v_a^2 m_a)$.
The axion energy density around the Earth is
\begin{align}
	\rho_a = \frac{1}{2}\dot a^2 + \frac{1}{2} m_a^2a^2 = \frac{1}{2} m_a^2 a_{\rm 0}^2 \simeq 0.3\,{\rm GeV/cm^3}.
\end{align}

The Lagrangian is
\begin{align}
	\mathcal L =-\frac{1}{4}F_{\mu\nu}F^{\mu\nu} - e A_\mu j^\mu 
	-\frac{1}{2}(\partial_\mu a)^2 - \frac{1}{2}m_a^2a^2 + \frac{a}{4M}F_{\mu\nu}\widetilde F^{\mu\nu},
\end{align}
where $\widetilde F^{\mu\nu}\equiv\epsilon^{\mu\nu\rho\sigma}F_{\rho\sigma}/2$ and the mass scale $M$ represents the strength of the axion-photon coupling. The equation of motion is given by
\begin{align}
	\Box A^\mu - \partial^\mu (\partial_\nu A^\nu)-\frac{1}{2M}\epsilon^{\mu\nu\rho\sigma}\partial_\nu(a F_{\rho\sigma}) = ej^\mu.
\end{align}
The conservation law of the current is satisfied: $\partial_\mu j^\mu=0$.
In components, the equation of motion leads to
\begin{align}
\begin{cases}
	&\mu=0~~:~~\Delta A^0 + \vec\nabla\cdot\dot{\vec A}=\frac{1}{M}(\vec\nabla a)\cdot\vec B + ej^0\\
	&\mu=i~~:~~\Box \vec A -\vec \nabla(\dot A^0 + \vec\nabla\cdot\vec A) =
	-\frac{1}{M}\left( \dot a\vec B + \nabla a\times \vec E \right) +e\vec j
\end{cases}
\end{align}
In terms of $\vec E$ and $\vec B$, it is written as
\begin{align}
\begin{cases}
	&\mu=0~~:~~-\vec\nabla\cdot\vec E=\frac{1}{M}(\vec\nabla a)\cdot\vec B + ej^0, \\
	&\mu=i~~:~~\dot{\vec E}-\vec\nabla\times\vec B=
	-\frac{1}{M}\left( \dot a\vec B + \nabla a\times \vec E \right) +e\vec j
\end{cases}
\label{nablaE_axion}
\end{align}
Assuming that the axion is almost spatially homogeneous $(|\dot a|\gg |\nabla a|)$ and the existence of background magnetic field $\vec B_0$ much larger than the axion-induced electric field, we obtain:
\begin{align}
	-\Box \vec E \simeq -\frac{\ddot a}{M}\vec B_0 + e\dot{\vec j}.
	\label{BoxE_axion}
\end{align}

\subsubsection{Electric field configuration}

Let us consider an effective one-dimensional setup in which the $x<0$ region is superconductor and $x>0$ region is vacuum. The uniform external magnetic field $\vec B_0$ is applied on the $z$ direction. The equation of motion of the electric field is
\begin{align}
	&-\Box \vec E \simeq -\frac{1}{M}\ddot a\vec B_0 & (x>0)\\
	&-\Box \vec E \simeq -\frac{1}{M}\ddot a\vec B_0 +e\dot {\vec j} & (x<0)
\end{align}
Here and in what follows we neglect the spatial dependence of the axion and just take $a(\vec x,t)=a_0\cos(m_a t)$. As for the electric current, we take $e\vec j=-\sigma \vec E$. The general solution is given by the sum of particular solution and the homogeneous solution:
\begin{align}
	&\vec E^{\rm (vac)} = \vec E_a^{\rm (vac)} + \vec E_h^{\rm (vac)} & (x>0) \label{Evac} \\
	&\vec E^{\rm (m)} = \vec E_a^{\rm (m)} + \vec E_h^{\rm (m)} & (x<0).
	\label{Em}
\end{align}
Here the homogeneous solution should satisfy
\begin{align}
	\ddot{\vec E}_h + \sigma \dot{\vec E}_h - \partial_x^2 \vec E_h = 0.
\end{align}

Below we solve the equation in complex form for convenience, since the equations are linear. The axion field is taken to be
\begin{align}
	a(t) = a_0 e^{im_a t}.
\end{align}
The particular solution that depends on the axion source term is given by
\begin{align}
	&\vec E_a^{\rm (vac)} = -\frac{a_0 B_0}{M}e^{im_a t}\vec e_z~~~~~~(x>0)\\
	&\vec E_a^{\rm (m)} = -\frac{m_a(m_a+i\sigma)}{m_a^2+\sigma^2}\frac{a_0B_0}{M}e^{im_at} \vec e_z~~~~~~(x<0).
\end{align}
The general homogeneous solutions are summation of arbitrary frequency modes $\omega$, but in the present situation it is sufficient to only consider the $\omega=m_a$ mode since otherwise the boundary conditions, explained below, will not be satisfied. We also only consider the out-going wave from the surface of the metal as a physically relevant situation. Thus the homogeneous solution is written as
\begin{align}
	&\vec E_h^{\rm (vac)} = E_1 e^{im_a(t-x)} \vec e_z~~~~~~~~~(x>0),\\
	&\vec E_h^{\rm (m)} = E_2 e^{i(m_at-k_R x)} e^{k_I x} \vec e_z~~~~~~~(x<0),
\end{align}
where $E_1$ and $E_2$ are constants, which will be determined by the boundary condition, and
\begin{align}
 	k_R= -\sqrt{\frac{m_a}{2}}\left[ 1+\sqrt{1+\left(\frac{\sigma}{m_a}\right)^2} \right]^{1/2},~~~~~~
	k_I= \sqrt{\frac{m_a}{2}}\left[ -1+\sqrt{1+\left(\frac{\sigma}{m_a}\right)^2} \right]^{1/2}.
\end{align}
The limiting form is given as
\begin{align}
	k_R\simeq \begin{cases}
		-m_a & (m_a\gg \sigma) \\
		-\sqrt{m_a\sigma/2} & (m_a\ll \sigma)
	\end{cases},~~~~~~
	k_I\simeq \begin{cases}
		\sigma & (m_a\gg \sigma) \\
		\sqrt{m_a\sigma/2} & (m_a\ll \sigma)
	\end{cases}.
\end{align}
Note that $\sqrt{m_a\sigma/2}^{-1} \equiv \delta$ is the skin depth of the metal (\ref{skin}).

In order to determine the coefficients $E_1$ and $E_2$ in the homogeneous solutions, we need to take the boundary condition at $x=0$ into account. The boundary conditions are given by
\begin{align}
	&E_z^{\rm (vac)} =E_z^{\rm (m)} ~~~{\rm at}~~~x=0,\\
	&B_y^{\rm (vac)} =B_y^{\rm (m)}~~~{\rm at}~~~x=0.
\end{align}
By using the Maxwell equation, it is also written as
\begin{align}
	&E_z^{\rm (vac)} =E_z^{\rm (m)} ~~~{\rm at}~~~x=0,\\
	&\partial_x E_z^{\rm (vac)} =\partial_x E_z^{\rm (m)}~~~{\rm at}~~~x=0.
\end{align}
By solving this, we obtain
\begin{align}
	&E_1=\frac{k_R+ik_I}{k_R-m_a+ik_I}\frac{\sigma(\sigma-im_a)}{m_a^2+\sigma^2}\frac{a_0B_0}{M},\\
	&E_2=\frac{m_a}{k_R-m_a+ik_I}\frac{\sigma(\sigma-im_a)}{m_a^2+\sigma^2}\frac{a_0B_0}{M}.
\end{align}

Let us take a limit $m_a\ll \sigma$. Then we have
\begin{align}
	&\vec E^{\rm (vac)} = \left[ \left(1-\frac{1+i}{2}\sqrt{\frac{2m_a}{\sigma}}\right)e^{im_a(t-x)}-e^{im_at}\right] \frac{a_0B_0}{M}\vec e_z,\\
	&\vec E^{\rm (m)} = \left[ -\frac{1+i}{2}\sqrt{\frac{2m_a}{\sigma}}e^{i(m_at-k_Rx)}e^{k_I x}-\frac{im_a}{\sigma}e^{im_at}\right] \frac{a_0B_0}{M}\vec e_z.
\end{align}
By taking only the real part, we obtain 
\begin{align}
	&\vec E^{\rm (vac)} = \left\{\cos(m_a(t-x))-\cos(m_at)+\sqrt{\frac{m_a}{2\sigma}}\left[ \sin(m_a(t-x))-\cos(m_a(t-x)) \right] \right\} \frac{a_0B_0}{M}\vec e_z,\\
	&\vec E^{\rm (m)} = \left\{ \sqrt{\frac{m_a}{2\sigma}}\left[ \sin\left(m_at-\frac{x}{\delta}\right)-\cos\left(m_at-\frac{x}{\delta}\right) \right]e^{x/\delta} 
	+\frac{m_a}{\sigma}\sin(m_at)\right\} \frac{a_0B_0}{M}\vec e_z.
\end{align}
This result at $x>0$ is the same as that of Ref.~\cite{Horns:2012jf} in the limit $\sigma\to\infty$. We found a correction due to the finite conductivity that is suppressed by $\sqrt{m_a/\sigma}$. It is usually negligible for a normal metal. The electric field at the surface of the metal $x=0$ is given as
\begin{align}
	 \vec E(x=0) =  \sqrt{\frac{m_a}{2\sigma}}\frac{a_0B_0}{M}\left[ \sin\left(m_at\right)-\cos\left(m_at\right) \right] \vec e_z.
	 \label{Ex0_metal_a}
\end{align}
It is smaller by a factor $\sim \sqrt{m_a/\sigma}$ than the estimate in Ref.~\cite{Iwazaki:2020zer}. The electric current is thus estimated as
\begin{align}
	-e\vec j(x=0) = \sigma\vec E(x=0)= \sqrt{\frac{m_a\sigma}{2}}\frac{a_0B_0}{M}\left[ \sin\left(m_at\right)-\cos\left(m_at\right) \right] \vec e_z.
\end{align}
It is also smaller by a factor $\sim \sqrt{m_a/\sigma}$ than the estimate in Ref.~\cite{Iwazaki:2020zer}. 
Assuming a cylindrical metal with radius $r$, the current is induced at its surface with a skin depth $\delta$. The total current is oscillating with time, and its amplitude is given by
\begin{align}
	I &= \sqrt{\frac{m_a\sigma}{2}}\frac{a_0B_0}{M} \times 2\pi r\delta\\
	&\simeq 1.1\times 10^{-16}\,{\rm A}~\left(\frac{B_0}{1\,{\rm T}}\right)\left(\frac{1\,{\rm \mu eV}}{m_a}\right)
	\left(\frac{10^{15}\,{\rm GeV}}{M}\right)\left(\frac{r}{1\,{\rm cm}}\right).
	\label{I_axion_metal}
\end{align}
Note that the dependence on the conductivity $\sigma$ is cancelled out.

\subsection{Hidden photon dark matter}  \label{sec:HPmetal}

\subsubsection{Basic equations}

Next we consider the hidden photon DM with the mass range of $\mu{\rm eV} \lesssim m_H\lesssim {\rm meV}$.  
As will soon be explained, the hidden photon field satisfies $\partial_\mu H^\mu=0$. We take the hidden photon DM field as
\begin{align}
	&H_0 (\vec x,t) \simeq -\vec v_H\cdot \vec H_{c}\cos\left( m_Ht - m_H \vec v_H\cdot\vec x \right),\\
	&\vec H (\vec x,t) \simeq \vec H_{c} \cos\left( m_Ht - m_H \vec v_H\cdot\vec x \right).
\end{align}
Thus the zeroth component $H_0$ is suppressed by the velocity $|\vec v_H|\sim 10^{-3}$ compared with the spatial component.
The hidden photon energy density around the Earth is
\begin{align}
	\rho_H \simeq \frac{1}{2}\dot{\vec H}^2 + \frac{1}{2} m_H^2{\vec H}^2 = \frac{1}{2} m_H^2 \vec{H}_c^2 \simeq 0.3\,{\rm GeV/cm^3}.
\end{align}

The Lagrangian is
\begin{align}
	\mathcal L =-\frac{1}{4}F_{\mu\nu}F^{\mu\nu} - e A_\mu j^\mu 
	-\frac{1}{4} H_{\mu\nu}H^{\mu\nu} - \frac{1}{2}m_H^2 H_\mu H^\mu  - \frac{\kappa}{2}F_{\mu\nu} H^{\mu\nu}.
\end{align}
where $H_{\mu\nu}\equiv \partial_\mu H_\nu-\partial_\nu H_\mu$ is the field strength tensor of the hidden photon and $\kappa$ denotes the kinetic mixing between the Standard Model photon and the hidden photon.
The equation of motion is given by
\begin{align}
	&\Box H^\mu - \partial^\mu (\partial_\nu H^\nu)-m_H^2 H^\mu + \kappa\left(\Box A^\mu - \partial^\mu (\partial_\nu A^\nu) \right)=0, \label{H_eom} \\
	&\Box A^\mu - \partial^\mu (\partial_\nu A^\nu)+\kappa\left(\Box H^\mu - \partial^\mu (\partial_\nu H^\nu) \right)= ej^\mu.
\end{align}
By multiplying $\partial_\mu$ on Eq.~(\ref{H_eom}), we find a constraint $\partial_\mu H^\mu = 0$. We also find a current conservation $\partial_\mu j^\mu = 0$. Then these equations become
\begin{align}
	&\Box H^\mu -m_H^2 H^\mu + \kappa\left(\Box A^\mu - \partial^\mu (\partial_\nu A^\nu) \right)=0, \\
	&\Box A^\mu - \partial^\mu (\partial_\nu A^\nu)+\kappa \Box H^\mu = ej^\mu.
\end{align}

In components, the equation of motion leads to
\begin{align}
\begin{cases}
	&\mu=0~~:~~\Delta A^0 + \vec\nabla\cdot\dot{\vec A}=-\kappa\Box H^0 + ej^0\\
	&\mu=i~~:~~\Box \vec A -\vec \nabla(\dot A^0 + \vec\nabla\cdot\vec A) =
	-\kappa\Box \vec H +e\vec j
\end{cases}
\end{align}
In terms of $\vec E$ and $\vec B$, it is written as
\begin{align}
\begin{cases}
	&\mu=0~~:~~-\vec\nabla\cdot\vec E=-\kappa\Box H^0 + ej^0\\
	&\mu=i~~:~~\dot{\vec E}-\vec\nabla\times\vec B=
	-\kappa\Box \vec H +e\vec j
\end{cases}
\label{nablaE_HP}
\end{align}
By neglecting the spatial dependence of $H_\mu$, we obtain
\begin{align}
	-\Box \vec E \simeq \kappa \frac{d^3}{dt^3}\vec H + e\dot{\vec j}.
	\label{BoxE_HP}
\end{align}

\subsubsection{Electric field configuration}

Comparing (\ref{BoxE_HP}) with (\ref{BoxE_axion}), the analysis will be almost parallel to the case of axion. The following replacement from the case of axion DM works: $m_a\to m_H$, $a_0\to H_c \cos\theta$, $B_0/M \to \kappa m_H$ and $\cos(m_at)\to \sin(m_H t)$ where $\theta$ is the angle between $\vec e_z$ and the vector $\vec H$. Note that the direction of $\vec H$ remains almost the same within one coherent time $\tau \sim 1/(v_H^2 m_H)$.
Contrary to the case of axion, we do not need to apply the external magnetic field in order to obtain nontrivial effects from the hidden photon DM.

In the limit $m_H\ll \sigma$, a parallel calculation to the axion case shows
\begin{align}
	&\vec E^{\rm (vac)} = \left\{\sin(m_H(t-x))-\sin(m_Ht)+\sqrt{\frac{m_H}{2\sigma}}\left[ \cos(m_H(t-x))-\sin(m_H(t-x)) \right] \right\} \kappa m_H H_c \cos\theta\vec e_z,\\
	&\vec E^{\rm (m)} = \left\{ \sqrt{\frac{m_H}{2\sigma}}\left[ \cos\left(m_Ht-\frac{x}{\delta}\right)-\sin\left(m_Ht-\frac{x}{\delta}\right) \right]e^{x/\delta} 
	+\frac{m_H}{\sigma}\cos(m_Ht)\right\}  \kappa m_H H_c \cos\theta \vec e_z.
\end{align}
The electric field at the surface of the metal $x=0$ is given as
\begin{align}
	 \vec E(x=0) =  \sqrt{\frac{m_H}{2\sigma}} \kappa m_H H_c \cos\theta \left[ \cos\left(m_Ht\right)-\sin\left(m_Ht\right) \right] \vec e_z.
	  \label{Ex0_metal_hp}
\end{align}
Note that this is smaller by a factor $\sim \sqrt{m_H/\sigma}$ than the electric field at the vacuum $(x\to \infty)$.
The electric current is thus estimated as
\begin{align}
	 -e\vec j(x=0)= \sigma\vec E(x=0) = \sqrt{\frac{m_H\sigma}{2}} \kappa m_H H_c \cos\theta \left[ \cos\left(m_Ht\right)-\sin\left(m_Ht\right) \right] \vec e_z.
\end{align}
Assuming a cylindrical metal with radius $r$, the amplitude of the total current is given by
\begin{align}
	I &= \sqrt{\frac{m_H\sigma}{2}}\kappa m_H H_c \cos\theta \times 2\pi r\delta\\
	&\simeq 5.5\times 10^{-16}\,{\rm A}~\left(\frac{\kappa}{10^{-15}}\right)\left(\frac{r}{1\,{\rm cm}}\right)\cos\theta.
	\label{I_HP_metal}
\end{align}

\section{Superconductor with dark matter background}  \label{sec:scdm}

In this section we consider the effect of axion or hidden photon DM on the electromagnetic configuration around the superconducting material. The most analysis is parallel to the case of metal performed in the previous section.
We assume that the DM mass is smaller than the energy gap of the superconducting material, which is typically $\mathcal O({\rm meV})$. Otherwise, the DM absorption destroys the Cooper pair and the excitation of quasi particles should be taken into account. The absorption of DM heavier than the superconductor gap has been considered in Refs.~\cite{Hochberg:2015pha,Hochberg:2016ajh}. In this sense, we consider the regime complementary to these previous studies.

\subsection{Axion dark matter} \label{sec:axionsc}

\subsubsection{Basic equations}

The Lagrangian of the electromagnetism in the superconductor, combined with the interaction with the axion, is given by
\begin{align}
	\mathcal L =-\frac{1}{4}F_{\mu\nu}F^{\mu\nu} - \frac{1}{2}m_A^2 A_\mu A^\mu 
	-\frac{1}{2}(\partial_\mu a)^2 - \frac{1}{2}m_a^2a^2 + \frac{a}{4M}F_{\mu\nu}\widetilde F^{\mu\nu}.
\end{align}
The equation of motion is given by
\begin{align}
	\Box A^\mu - \partial^\mu (\partial_\nu A^\nu)-\frac{1}{2M}\epsilon^{\mu\nu\rho\sigma}\partial_\nu(a F_{\rho\sigma}) = ej_A^\mu,
\end{align}
where $ej_A^\mu = m_A^2 A^\mu$. In components, the equation of motion leads to
\begin{align}
\begin{cases}
	&\mu=0~~:~~\Delta A^0 + \vec\nabla\cdot\dot{\vec A}=\frac{1}{M}(\vec\nabla a)\cdot\vec B + ej_A^0\\
	&\mu=i~~:~~\Box \vec A -\vec \nabla(\dot A^0 + \vec\nabla\cdot\vec A) =
	-\frac{1}{M}\left( \dot a\vec B + \nabla a\times \vec E \right) +e\vec j_A
\end{cases}
\end{align}
In terms of $\vec E$ and $\vec B$, it is written as
\begin{align}
\begin{cases}
	&\mu=0~~:~~-\vec\nabla\cdot\vec E=\frac{1}{M}(\vec\nabla a)\cdot\vec B + ej_A^0, \\
	&\mu=i~~:~~\dot{\vec E}-\vec\nabla\times\vec B=
	-\frac{1}{M}\left( \dot a\vec B + \nabla a\times \vec E \right) +e\vec j_A.
\end{cases}
\end{align}
Assuming spatially homogeneous axion field, it leads to
\begin{align}
	 &(-\Box +m_A^2) \vec E \simeq -\frac{1}{M}\ddot a\vec B_0, \\
	 &(-\Box +m_A^2) \vec B \simeq \frac{1}{M}\dot a \vec\nabla\times\vec B.
\end{align}

\subsubsection{Electric field configuration}

Let us consider an effective one-dimensional setup in which the $x<0$ region is superconductor and $x>0$ region is vacuum. Let us apply the external magnetic field $\vec B_0$ along the $z$ direction. It is screened inside the superconductor, so the static magnetic field configuration is given as
\begin{align}
	&\vec B_0(x) = 
	\begin{cases}
	B_0 \vec e_z ~~~~~&(x>0)\\
	B_0 \exp(m_A x) \vec e_z  ~~~~~&(x<0)
	\end{cases}.
\end{align}
The equation of motion of the electric field is
\begin{align}
	&-\Box \vec E = -\frac{1}{M}\ddot a\vec B_0 & (x>0)\\
	&(-\Box +m_A^2) \vec E = -\frac{1}{M}\ddot a\vec B_0 & (x<0).
\end{align}
Here and in what follows we neglect the spatial dependence of the axion and just take $a(\vec x,t)=a_0\cos(m_a t)$.
The general solution is given by the sum of particular solution and the homogeneous solution:
\begin{align}
	&\vec E^{\rm (vac)} = \vec E_a^{\rm (vac)} + \vec E_h^{\rm (vac)} & (x>0) \label{Evac} \\
	&\vec E^{\rm (sc)} = \vec E_a^{\rm (sc)} + \vec E_h^{\rm (sc)} & (x<0).
	\label{Esc}
\end{align}

The particular solution is given by
\begin{align}
	&\vec E_a^{\rm (vac)} = -\frac{a_0B_0}{M}\cos(m_a t)\vec e_z,\\
	&\vec E_a^{\rm (sc)} = -\frac{a_0B_0 e^{m_Ax}}{M}\cos(m_at) \vec e_z.
\end{align}
The general homogeneous solutions are given by
\begin{align}
	&\vec E_h^{\rm (vac)} = E_1' \vec e_z\cos(m_a(t-x)) +  E_1'' \vec e_z \sin(m_a(t-x)) , \\
	&\vec E_h^{\rm (sc)} = \left(E_2'\cos(m_at) + E_2''\sin(m_at) \right)\vec e_z e^{\sqrt{m_A^2-m_a^2}x}.
\end{align}
To determine the coefficients $E_1', E_1'', E_2'$ and $E_2''$ we need to take account of the boundary condition at the boundary $x=0$.
The conditions are
\begin{align}
	&E_z^{\rm (vac)} =E_z^{\rm (sc)} ~~~{\rm at}~~~x=0,\\
	&\partial_x E_z^{\rm (vac)} =\partial_x E_z^{\rm (sc)}~~~{\rm at}~~~x=0.
\end{align}
By solving this, we obtain
\begin{align}
	&E_1' = E_2' = \frac{\sqrt{m_A^2-m_a^2}}{m_A}\frac{a_0 B_0}{M},~~~~~~E_1''=E_2''= \frac{m_a}{m_A}\frac{a_0 B_0}{M}. 
\end{align}
Substituting this into Eqs.~(\ref{Evac},\ref{Esc}), we finally obtain the electric field configuration as
\begin{align}
	&\vec E^{\rm (vac)} =\frac{a_0B_0}{M}\vec e_z\left[-\cos(m_at)+\frac{\sqrt{m_A^2-m_a^2}}{m_A}\cos(m_a(t-x))+\frac{m_a}{m_A}\sin(m_a(t-x))\right],\\
	&\vec E^{\rm (sc)} =\frac{a_0B_0}{M}\vec e_z
	\left[-\cos(m_at)e^{m_Ax} +\left(\frac{\sqrt{m_A^2-m_a^2}}{m_A}\cos(m_at)+\frac{m_a}{m_A}\sin(m_at)\right)e^{\sqrt{m_A^2-m_a^2}x} \right].
\end{align}
In particular, assuming $m_a\ll m_A$, the electric field at the boundary $x=0$ is
\begin{align}
	\vec E(x=0)\simeq \frac{a_0B_0}{M}\frac{m_a}{m_A} \sin(m_at) \vec e_z,
\end{align}
It is smaller by a factor $\sim (m_a/m_A)$ than the estimate of Refs.~\cite{Iwazaki:2020agl,Iwazaki:2020zer}.
The current at the surface $x\simeq 0$ is calculated from the London equation (\ref{London}):\footnote{
	The same result is obtained by a phenomenological estimate: $e\vec j \sim en_e \vec v_e$ where $\vec v_e \sim e\vec E/(m_em_a)$, and hence $e\vec j \sim m_A^2\vec E/m_a$.
}
\begin{align}
	e\vec j_A\simeq \frac{a_0 B_0 m_A}{M}\cos(m_a t) \vec e_z.
\end{align}
This is also smaller by a factor $\sim (m_a/m_A)$ than the estimate of Refs.~\cite{Iwazaki:2020agl,Iwazaki:2020zer}.
Assuming a cylindrical superconductor with radius $r$, the current is induced at its surface with a thin layer of $m_A^{-1}$. The total current is oscillating with time, and its amplitude is given by
\begin{align}
	I &= \frac{a_0B_0 m_A}{M} \times 2\pi r m_A^{-1}\\
	&\simeq 1.1\times 10^{-16}\,{\rm A}~\left(\frac{B_0}{1\,{\rm T}}\right)\left(\frac{1\,{\rm \mu eV}}{m_a}\right)
	\left(\frac{10^{15}\,{\rm GeV}}{M}\right)\left(\frac{r}{1\,{\rm cm}}\right).
\end{align}
It is the same expression as the case of metal (\ref{I_axion_metal}).

\subsection{Hidden photon dark matter} \label{sec:HPsc}

\subsubsection{Basic equations}

Next we discuss the case of hidden photon DM with a superconductor. The Lagrangian is
\begin{align}
	\mathcal L =-\frac{1}{4}F_{\mu\nu}F^{\mu\nu} -\frac{1}{2}m_A^2 A_\mu A^\mu 
	-\frac{1}{4} H_{\mu\nu}H^{\mu\nu} - \frac{1}{2}m_H^2 H_\mu H^\mu  - \frac{\kappa}{2}F_{\mu\nu} H^{\mu\nu}.
\end{align}
The equation of motions are similar to the case of metal discussed in Sec.~\ref{sec:HPmetal}. What we need is only to replace $ej^\mu\to ej_A^\mu=m_A^2A^\mu$. The equations are
\begin{align}
\begin{cases}
	&\mu=0~~:~~\Delta A^0 + \vec\nabla\cdot\dot{\vec A}=-\kappa\Box H^0 + ej_A^0\\
	&\mu=i~~:~~\Box \vec A -\vec \nabla(\dot A^0 + \vec\nabla\cdot\vec A) =
	-\kappa\Box \vec H +e\vec j
\end{cases}
\end{align}
In terms of $\vec E$ and $\vec B$, it is written as
\begin{align}
\begin{cases}
	&\mu=0~~:~~-\vec\nabla\cdot\vec E=-\kappa\Box H^0 + ej_A^0\\
	&\mu=i~~:~~\dot{\vec E}-\vec\nabla\times\vec B=
	-\kappa\Box \vec H +e\vec j_A
\end{cases}
\end{align}
Assuming the spatially homogeneous hidden photon DM, we can derive the equation of the electric field as
\begin{align}
	\left(-\Box + m_A^2\right)\vec E \simeq \kappa \frac{d^3}{dt^3}\vec H.
\end{align}

\subsubsection{Electric field configuration}

The equation of motion of the electric field is
\begin{align}
	&-\Box \vec E = \kappa \frac{d^3}{dt^3}\vec H & (x>0)\\
	&(-\Box +m_A^2) \vec E = \kappa \frac{d^3}{dt^3}\vec H & (x<0).
\end{align}
The general solution is given by the sum of particular solution and the homogeneous solution:
\begin{align}
	&\vec E^{\rm (vac)} = \vec E_H^{\rm (vac)} + \vec E_h^{\rm (vac)} & (x>0)\\
	&\vec E^{\rm (sc)} = \vec E_H^{\rm (sc)} + \vec E_h^{\rm (sc)} & (x<0).
\end{align}
The particular solution is given by
\begin{align}
	&\vec E_H^{\rm (vac)} = -\kappa m_H H_c\cos\theta \sin(m_H t)\vec e_z,\\
	&\vec E_H^{\rm (sc)} = -\frac{m_H^2}{m_H^2-m_A^2}\kappa m_H H_c\cos\theta \sin(m_Ht) \vec e_z.
\end{align}
The general homogeneous solutions are the same as the case of superconductor: it is given by
\begin{align}
	&\vec E_h^{\rm (vac)} = E_1' \vec e_z\cos(m_H(t-x)) +  E_1'' \vec e_z \sin(m_H(t-x)) , \\
	&\vec E_h^{\rm (sc)} = \left(E_2'\cos(m_Ht) + E_2''\sin(m_Ht) \right)\vec e_z e^{\sqrt{m_A^2-m_H^2}x}.
\end{align}
To determine the coefficients $E_1', E_1'', E_2'$ and $E_2''$ we need to take account of the boundary condition at the boundary $x=0$. 
Repeating the same procedure as the previous section, we obtain
\begin{align}
	&E_1'=E_2'=-\frac{m_H}{\sqrt{m_A^2-m_H^2}}\kappa m_H H_c\cos\theta,\\
	&E_1''=\kappa m_H H_c\cos\theta,~~~~~~E_2''=-\frac{m_H^2}{m_A^2-m_H^2}\kappa m_H H_c\cos\theta.
\end{align}
Thus the full solution is
\begin{align}
	&\vec E^{\rm (vac)} = \left[-\sin(m_Ht)+\sin\left(m_H(t-x)\right)-\frac{m_H}{\sqrt{m_A^2-m_H^2}}\cos\left(m_H(t-x)\right) \right] \kappa m_H H_c\cos\theta\,\vec e_z,\\
	&\vec E^{\rm (sc)} = \left\{\frac{m_H^2}{m_A^2-m_H^2}\sin(m_Ht)-\frac{m_H}{\sqrt{m_A^2-m_H^2}}\left[\cos(m_Ht) +\frac{m_H}{\sqrt{m_A^2-m_H^2}}\sin(m_Ht) \right] e^{\sqrt{m_A^2-m_H^2}x} \right\} \nonumber\\
	&~~~~~~~~~~~~\times \kappa m_H H_c\cos\theta\,\vec e_z
\end{align}
Therefore, at the surface of the superconductor $x=0$, the electric field is given by
\begin{align}
	\vec E(x=0)=-\frac{m_H}{\sqrt{m_A^2-m_H^2}}\kappa m_H H_c\cos\theta \cos\left(m_H t\right) \vec e_z.
\end{align}
It is smaller by a factor $\sim m_H/m_A$ than the value at the vacuum $(x\to \infty)$. The electric current is estimated as
\begin{align}
	e\vec j_A \simeq -\kappa m_A m_H H_c\cos\theta \sin\left(m_H t\right) \vec e_z.
\end{align}
Assuming a cylindrical superconductor with radius $r$, the amplitude of the total current is given by
\begin{align}
	I &=\kappa m_A m_H H_c \cos\theta \times 2\pi r m_A^{-1}\\
	&\simeq 5.5\times 10^{-16}\,{\rm A}~\left(\frac{\kappa}{10^{-15}}\right)\left(\frac{r}{1\,{\rm cm}}\right)\cos\theta.
\end{align}
This is the same expression as the case of metal (\ref{I_HP_metal}).

\section{Detection possibility} \label{sec:det}

In the previous sections we derived surface electric current on a material induced by axion/hidden-photon DM. Now let us discuss a possibility to measure such a current. Assuming a cylindrical metal, the total power compensated as the Joule heat is given by
\begin{align}
	P_{\rm J} = \int dV \sigma \vec E^2(x,t) \sim \sigma \vec E^2(x=0,t) \Delta V,
	\label{PJ}
\end{align}
where $\Delta V \simeq 2\pi r\ell\delta$ with $r$, $\ell$ and $\delta$ being the radius, height and the skin depth of the material, respectively. 
Substituting $\vec E(x=0,t)$ given in Eqs.~(\ref{Ex0_metal_a}) for the axion DM and (\ref{Ex0_metal_hp}) for the hidden photon DM and taking the time average over the DM oscillation time scale, we obtain
\begin{align}
	P_{\rm J} = \begin{cases}
		2m_{a} \rho_{a} \Delta V \left(\frac{B_0}{m_a M}\right)^2& {\rm for~axion}\\
	 	2m_{H} \rho_{H} \Delta V (\kappa \cos\theta)^2 &  {\rm for~hidden~photon}
	\end{cases}.
\end{align}
On the other hand, the Johnson-Nyquist noise power at the temperature $T$ and the frequency range $[\omega,\omega+\Delta\omega]$ is given by
\begin{align}
	P_{\rm T} = \frac{T \Delta\omega}{2\pi}.
\end{align}
Taking the ratio, we obtain
\begin{align}
	\frac{P_{\rm J}}{P_{\rm T}} \simeq 6\times 10^{-11}\left(\frac{\ell}{1\,{\rm m}}\right)\left(\frac{r}{1\,{\rm cm}}\right)
	\left(\frac{1\,{\rm K}}{T}\right)\left(\frac{10^4\,{\rm eV}}{\sigma}\right)^{\frac{1}{2}}\left(\frac{1\,{\rm \mu eV}}{m_{a}}\right)^{\frac{5}{2}}
	\left(\frac{10^{15}\,{\rm GeV}}{M}\right)^2\left(\frac{B_0}{10\,{\rm T}}\right)^2\left(\frac{\mathcal A}{10^6}\right),
\end{align}
for the axion DM, and
\begin{align}
	\frac{P_{\rm J}}{P_{\rm T}} \simeq 2\times 10^{-11}\left(\frac{\ell}{1\,{\rm m}}\right)\left(\frac{r}{1\,{\rm cm}}\right)
	\left(\frac{1\,{\rm K}}{T}\right)\left(\frac{10^4\,{\rm eV}}{\sigma}\right)^{\frac{1}{2}}\left(\frac{1\,{\rm \mu eV}}{m_{H}}\right)^{\frac{1}{2}}
	\left(\frac{\kappa\cos\theta}{10^{-15}}\right)^2\left(\frac{\mathcal A}{10^6}\right),
\end{align}
for the hidden photon DM, where we defined $\mathcal A\equiv m_{a/H}/(2\Delta\omega)$.

The DM signal may be significantly enhanced if we use the LC circuit to detect the current on a metal. Let us connect the top and bottom of the cylindrical material with a conducting wire, which then forms a LC circuit. The cylindrical material under the DM background acts as an alternating current source of the circuit. In the LC circuit, there is a resonant frequency at $\omega = \omega_0\equiv 1/\sqrt{LC}$ where $L$ is the inductance and $C$ is the capacitance. At $m_{a/H}=\omega_0$, the signal is enhanced by the $Q$-value of the circuit, where $Q=\omega_0/(2\Delta\omega) = (1/R)\sqrt{L/C}$ with $R$ being the resistance and $\Delta\omega =R/(2L)$ the resonance width in a series LC circuit. The energy acquired by the coil is given by $U_L= LI^2(t)/2$ and hence $\dot U_L \sim \omega_0 L I_0^2 \sim Q R I_0^2$ where $I_0$ is the amplitude of the oscillating current. It is larger by a factor $Q$ than the power (\ref{PJ}).
Thus we may improve the sensitivity by measuring the oscillating current or the magnetic field in the coil using the resonant LC circuit.
Note that the time-averaged work done by the electromotive force, which is the DM-induced electric field, is given by $(1/T)\int_0^{T} \ell E(t) I(t)dt \sim \omega_0 L I_0^2/Q \sim R I_0^2$ and it is the same as (\ref{PJ}).\footnote{
    Note that the resistance is given by $R \simeq \ell/(2\pi\delta r\sigma)$.
} The signal-to-noise ratio is then given by
\begin{align}
\frac{S}{N} = \frac{P_{\rm J}}{P_{\rm T}}Q\sqrt{\frac{\Delta\omega \,t_{\rm obs}}{2\pi}}
\simeq 9\times 10^7 \left(\frac{Q}{10^6}\right)^{\frac{1}{2}}
\left(\frac{m_{a/H}}{1\,{\rm \mu eV}}\right)^{\frac{1}{2}}
\left(\frac{t_{\rm obs}}{1\,{\rm min}}\right)^{\frac{1}{2}}
\frac{P_{\rm J}}{P_{\rm T}},
\end{align}
where $t_{\rm obs}$ is the observation time at each band. Thus it is not very unrealistic that this method covers the QCD axion parameter region.\footnote{
    In the case of QCD axion, the axion-photon coupling $M$ is related to the axion mass $m_a$ through $M = \mathcal C \times (5\times 10^{15}\,{\rm GeV})\times(1\,{\rm \mu eV}/m_a)$ with $\mathcal C$ being $\mathcal O(1)$ coefficients depending on the detail of the axion model.
}
This method can be effective when the coupling to photons is stronger.
In particular, it may have an advantage in the case where the Haloscope is not realistic because the mass is too small and the cavity size is too large, or when the mass is large and the cavity size and $Q$ value are small.

\section{Summary} \label{sec:dis}

We derived the electric field configuration and the electric current around a surface of metal or superconductor under the axion/hidden-photon DM background. The current we found is orders-of-magnitude smaller than the estimates in Refs.~\cite{Iwazaki:2020agl,Iwazaki:2020zer}, although still it may be relevant for future/on-going experiments. For example, in the case of experiments using mirrors such as MADMAX or BRASS~\cite{Horns:2012jf,TheMADMAXWorkingGroup:2016hpc}, the electric field far from the mirror surface is almost identical to the previous estimates but there is a small correction suppressed by $\sqrt{m/\sigma}$ with $m$ and $\sigma$ being the DM mass and the conductivity, respectively. A similar correction is also expected in a cavity setup such as ADMX~\cite{Asztalos:2009yp}, HAYSTAC~\cite{Zhong:2018rsr}, ORGAN~\cite{McAllister:2017lkb}, KLASH~\cite{Alesini:2017ifp} and CULTASK~\cite{Semertzidis:2019gkj}, although such an effect is absorbed in the cavity quality factor $Q$ and it does not give a significant correction.
We also briefly discussed a possibility to detect such an oscillating electric current by using the LC circuit.

\section*{Acknowledgments}
This work was supported by JSPS KAKENHI Grant (Nos. 21K18621 [YK], 21H05446 [YK], 18K03609 [KN], 17H06359 [KN]).




\begin{thebibliography}{99}


\bibitem{Preskill:1982cy}
  J.~Preskill, M.~B.~Wise and F.~Wilczek,
  Phys.\ Lett.\ B {\bf 120}, 127 (1983).

\bibitem{Abbott:1982af}
  L.~F.~Abbott and P.~Sikivie,
  Phys.\ Lett.\ B {\bf 120}, 133 (1983).

\bibitem{Dine:1982ah}
  M.~Dine and W.~Fischler,
  Phys.\ Lett.\ B {\bf 120}, 137 (1983).

\bibitem{Kim:1986ax}
  J.~E.~Kim,
  Phys.\ Rept.\  {\bf 150}, 1 (1987).

\bibitem{Kawasaki:2013ae}
  M.~Kawasaki and K.~Nakayama,
  Ann.\ Rev.\ Nucl.\ Part.\ Sci.\  {\bf 63}, 69 (2013)
  [arXiv:1301.1123 [hep-ph]].







\bibitem{Agrawal:2018vin}
  P.~Agrawal, N.~Kitajima, M.~Reece, T.~Sekiguchi and F.~Takahashi,
  arXiv:1810.07188 [hep-ph].

\bibitem{Co:2018lka}
  R.~T.~Co, A.~Pierce, Z.~Zhang and Y.~Zhao,
  Phys.\ Rev.\ D {\bf 99}, no. 7, 075002 (2019)
  [arXiv:1810.07196 [hep-ph]].

\bibitem{Bastero-Gil:2018uel}
  M.~Bastero-Gil, J.~Santiago, L.~Ubaldi and R.~Vega-Morales,
  JCAP {\bf 1904}, no. 04, 015 (2019)
  [arXiv:1810.07208 [hep-ph]].

\bibitem{Dror:2018pdh}
  J.~A.~Dror, K.~Harigaya and V.~Narayan,
  Phys.\ Rev.\ D {\bf 99}, no. 3, 035036 (2019)
  [arXiv:1810.07195 [hep-ph]].

\bibitem{Long:2019lwl}
  A.~J.~Long and L.~T.~Wang,
  Phys.\ Rev.\ D {\bf 99}, no. 6, 063529 (2019)
  [arXiv:1901.03312 [hep-ph]].

\bibitem{Graham:2015rva}
  P.~W.~Graham, J.~Mardon and S.~Rajendran,
  Phys.\ Rev.\ D {\bf 93}, no. 10, 103520 (2016)
  [arXiv:1504.02102 [hep-ph]].

\bibitem{Ema:2019yrd}
  Y.~Ema, K.~Nakayama and Y.~Tang,
  JHEP {\bf 1907}, 060 (2019)
  [arXiv:1903.10973 [hep-ph]].
  
\bibitem{Ahmed:2020fhc}
A.~Ahmed, B.~Grzadkowski and A.~Socha,
JHEP \textbf{08}, 059 (2020)
[arXiv:2005.01766 [hep-ph]].

\bibitem{Kolb:2020fwh}
E.~W.~Kolb and A.~J.~Long,
JHEP \textbf{03}, 283 (2021)
[arXiv:2009.03828 [astro-ph.CO]].

\bibitem{Nakayama:2021avl}
K.~Nakayama and W.~Yin,
JHEP \textbf{10}, 026 (2021)
[arXiv:2105.14549 [hep-ph]].




\bibitem{Sikivie:1983ip}
  P.~Sikivie,
  Phys.\ Rev.\ Lett.\  {\bf 51}, 1415 (1983)
  Erratum: [Phys.\ Rev.\ Lett.\  {\bf 52}, 695 (1984)].

\bibitem{Bradley:2003kg}
  R.~Bradley, J.~Clarke, D.~Kinion, L.~J.~Rosenberg, K.~van Bibber, S.~Matsuki, M.~Muck and P.~Sikivie,
  Rev.\ Mod.\ Phys.\  {\bf 75}, 777 (2003).

\bibitem{Asztalos:2009yp}
  S.~J.~Asztalos {\it et al.} [ADMX Collaboration],
  Phys.\ Rev.\ Lett.\  {\bf 104}, 041301 (2010)
  [arXiv:0910.5914 [astro-ph.CO]].

\bibitem{Zhong:2018rsr}
  L.~Zhong {\it et al.} [HAYSTAC Collaboration],
  Phys.\ Rev.\ D {\bf 97}, no. 9, 092001 (2018)
  [arXiv:1803.03690 [hep-ex]].

\bibitem{McAllister:2017lkb}
  B.~T.~McAllister, G.~Flower, E.~N.~Ivanov, M.~Goryachev, J.~Bourhill and M.~E.~Tobar,
  Phys.\ Dark Univ.\  {\bf 18}, 67 (2017)
  [arXiv:1706.00209 [physics.ins-det]].

\bibitem{Alesini:2017ifp}
  D.~Alesini, D.~Babusci, D.~Di Gioacchino, C.~Gatti, G.~Lamanna and C.~Ligi,
  arXiv:1707.06010 [physics.ins-det].

\bibitem{Semertzidis:2019gkj}
  Y.~K.~Semertzidis {\it et al.},
  arXiv:1910.11591 [physics.ins-det].

\bibitem{Horns:2012jf}
  D.~Horns, J.~Jaeckel, A.~Lindner, A.~Lobanov, J.~Redondo and A.~Ringwald,
  JCAP {\bf 1304}, 016 (2013)
  [arXiv:1212.2970 [hep-ph]].

\bibitem{Jaeckel:2013eha}
  J.~Jaeckel and J.~Redondo,
  Phys.\ Rev.\ D {\bf 88}, no. 11, 115002 (2013)
  [arXiv:1308.1103 [hep-ph]].

\bibitem{TheMADMAXWorkingGroup:2016hpc}
  A.~Caldwell {\it et al.} [MADMAX Working Group],
  Phys.\ Rev.\ Lett.\  {\bf 118}, no. 9, 091801 (2017)
  [arXiv:1611.05865 [physics.ins-det]].




\bibitem{Sikivie:2013laa}
P.~Sikivie, N.~Sullivan and D.~B.~Tanner,
Phys. Rev. Lett. \textbf{112}, no.13, 131301 (2014)
[arXiv:1310.8545 [hep-ph]].

\bibitem{Chaudhuri:2014dla}
S.~Chaudhuri, P.~W.~Graham, K.~Irwin, J.~Mardon, S.~Rajendran and Y.~Zhao,
Phys. Rev. D \textbf{92}, no.7, 075012 (2015)
[arXiv:1411.7382 [hep-ph]].

\bibitem{Kahn:2016aff}
  Y.~Kahn, B.~R.~Safdi and J.~Thaler,
  Phys.\ Rev.\ Lett.\  {\bf 117}, no. 14, 141801 (2016)
  [arXiv:1602.01086 [hep-ph]].
  
  
  

\bibitem{Obata:2018vvr}
  I.~Obata, T.~Fujita and Y.~Michimura,
  Phys.\ Rev.\ Lett.\  {\bf 121}, no. 16, 161301 (2018)
  [arXiv:1805.11753 [astro-ph.CO]].

\bibitem{Nagano:2019rbw}
  K.~Nagano, T.~Fujita, Y.~Michimura and I.~Obata,
  Phys.\ Rev.\ Lett.\  {\bf 123}, no. 11, 111301 (2019)
  [arXiv:1903.02017 [hep-ph]].

\bibitem{Lawson:2019brd}
  M.~Lawson, A.~J.~Millar, M.~Pancaldi, E.~Vitagliano and F.~Wilczek,
  Phys.\ Rev.\ Lett.\  {\bf 123} (2019) no.14,  141802
  [arXiv:1904.11872 [hep-ph]].

\bibitem{Zarei:2019sva}
  M.~Zarei, S.~Shakeri, M.~Abdi, D.~J.~E.~Marsh and S.~Matarrese,
  arXiv:1910.09973 [hep-ph].

\bibitem{Budker:2013hfa}
  D.~Budker, P.~W.~Graham, M.~Ledbetter, S.~Rajendran and A.~Sushkov,
  Phys.\ Rev.\ X {\bf 4}, no. 2, 021030 (2014)
  [arXiv:1306.6089 [hep-ph]].

\bibitem{Barbieri:1985cp}
  R.~Barbieri, M.~Cerdonio, G.~Fiorentini and S.~Vitale,
  Phys.\ Lett.\ B {\bf 226}, 357 (1989).

\bibitem{Barbieri:2016vwg}
  R.~Barbieri {\it et al.},
  Phys.\ Dark Univ.\  {\bf 15}, 135 (2017)
  [arXiv:1606.02201 [hep-ph]].

\bibitem{Chigusa:2020gfs}
S.~Chigusa, T.~Moroi and K.~Nakayama,
Phys. Rev. D \textbf{101}, no.9, 096013 (2020)
[arXiv:2001.10666 [hep-ph]].

\bibitem{Ikeda:2021mlv}
T.~Ikeda, A.~Ito, K.~Miuchi, J.~Soda, H.~Kurashige and Y.~Shikano,
[arXiv:2102.08764 [hep-ex]].

\bibitem{Marsh:2018dlj}
  D.~J.~E.~Marsh, K.~C.~Fong, E.~W.~Lentz, L.~Smejkal and M.~N.~Ali,
  Phys.\ Rev.\ Lett.\  {\bf 123}, no. 12, 121601 (2019)
  [arXiv:1807.08810 [hep-ph]].

\bibitem{Schutte-Engel:2021bqm}
J.~Sch\"utte-Engel, D.~J.~E.~Marsh, A.~J.~Millar, A.~Sekine, F.~Chadha-Day, S.~Hoof, M.~N.~Ali, K.~C.~Fong, E.~Hardy and L.~\v{S}mejkal,
JCAP \textbf{08}, 066 (2021)
[arXiv:2102.05366 [hep-ph]].

\bibitem{Chigusa:2021mci}
S.~Chigusa, T.~Moroi and K.~Nakayama, 
JHEP \textbf{08}, 074 (2021)
[arXiv:2102.06179 [hep-ph]].


\bibitem{Crescini:2020cvl}
  N.~Crescini {\it et al.},
  arXiv:2001.08940 [hep-ex].
  
\bibitem{Hochberg:2015pha}
Y.~Hochberg, Y.~Zhao and K.~M.~Zurek,
Phys. Rev. Lett. \textbf{116}, no.1, 011301 (2016)
[arXiv:1504.07237 [hep-ph]].

\bibitem{Hochberg:2016ajh}
Y.~Hochberg, T.~Lin and K.~M.~Zurek,
Phys. Rev. D \textbf{94}, no.1, 015019 (2016)
[arXiv:1604.06800 [hep-ph]].
 
\bibitem{Hochberg:2016sqx}
Y.~Hochberg, T.~Lin and K.~M.~Zurek,
Phys. Rev. D \textbf{95}, no.2, 023013 (2017)
[arXiv:1608.01994 [hep-ph]].

\bibitem{Hochberg:2017wce}
Y.~Hochberg, Y.~Kahn, M.~Lisanti, K.~M.~Zurek, A.~G.~Grushin, R.~Ilan, S.~M.~Griffin, Z.~F.~Liu, S.~F.~Weber and J.~B.~Neaton,
Phys. Rev. D \textbf{97}, no.1, 015004 (2018)
[arXiv:1708.08929 [hep-ph]].

\bibitem{Knapen:2017ekk}
S.~Knapen, T.~Lin, M.~Pyle and K.~M.~Zurek,
Phys. Lett. B \textbf{785}, 386-390 (2018)
[arXiv:1712.06598 [hep-ph]].

\bibitem{Griffin:2018bjn}
S.~Griffin, S.~Knapen, T.~Lin and K.~M.~Zurek,
Phys. Rev. D \textbf{98}, no.11, 115034 (2018)
[arXiv:1807.10291 [hep-ph]].

\bibitem{Hochberg:2019cyy}
Y.~Hochberg, I.~Charaev, S.~W.~Nam, V.~Verma, M.~Colangelo and K.~K.~Berggren,
Phys. Rev. Lett. \textbf{123}, no.15, 151802 (2019)
[arXiv:1903.05101 [hep-ph]].

\bibitem{Campbell-Deem:2019hdx}
B.~Campbell-Deem, P.~Cox, S.~Knapen, T.~Lin and T.~Melia,
Phys. Rev. D \textbf{101}, no.3, 036006 (2020)
[erratum: Phys. Rev. D \textbf{102}, no.1, 019904 (2020)]
[arXiv:1911.03482 [hep-ph]].

\bibitem{Mitridate:2020kly}
A.~Mitridate, T.~Trickle, Z.~Zhang and K.~M.~Zurek,
Phys. Rev. D \textbf{102}, no.9, 095005 (2020)
[arXiv:2005.10256 [hep-ph]].

\bibitem{Hochberg:2021yud}
Y.~Hochberg, B.~V.~Lehmann, I.~Charaev, J.~Chiles, S.~W.~Nam and K.~K.~Berggren,
[arXiv:2110.01586 [hep-ph]].
  

\bibitem{Iwazaki:2020agl}
A.~Iwazaki,
Phys. Lett. B \textbf{811}, 135861 (2020)
[arXiv:2007.09832 [hep-ph]].

\bibitem{Iwazaki:2020zer}
A.~Iwazaki,
Nucl. Phys. B \textbf{963}, 115298 (2021)
[arXiv:2009.12212 [hep-ph]].




\end{thebibliography}
\end{document}